\documentstyle[prd,preprint,tighten,aps,eqsecnum,amssymb,epsfig]{revtex}
\begin{document}
\draft
\preprint{
\begin{tabular}{r}
DFTT 2/96
\\
JHU-TIPAC 96002
\\
hep-ph/9602216
\end{tabular}
}
\title{Short-baseline
neutrino oscillations
and neutrinoless double-beta decay
in the framework
of three neutrino mixing and a mass hierarchy}
\author{
S.M. Bilenky$^{\mathrm{a}}$,
A. Bottino$^{\mathrm{b}}$,
C. Giunti$^{\mathrm{b}}$
and
C. W. Kim$^{\mathrm{c}}$
}
\address{
\begin{tabular}{c}
$^{\mathrm{a}}$Joint Institute for Nuclear Research,
Dubna, Russia,
and
\\
SISSA-ISAS, Trieste, Italy.
\\
$^{\mathrm{b}}$INFN,
Sezione di Torino and Dipartimento di Fisica Teorica,
Universit\`a di Torino,
\\
Via P. Giuria 1, 10125 Torino, Italy.
\\
$^{\mathrm{c}}$Department of Physics and Astronomy,
The Johns Hopkins University,
\\
Baltimore, Maryland 21218, USA.
\end{tabular}
}
\date{February 2, 1996}
\maketitle
\begin{abstract}
We have analyzed
the results of
the latest terrestrial neutrino oscillation experiments
in the framework of a model with
mixing of three massive neutrinos
and a neutrino mass hierarchy
($ m_1 \ll m_2 \ll m_3 $).
In this model,
oscillations of the terrestrial neutrinos
are characterized by three parameters,
$ \Delta m^2 \equiv m_3^2 - m_1^2 $
and the squared moduli of
the two mixing matrix elements
$U_{e3}$ and $U_{\mu3}$.
Using the results of disappearance experiments
and solar neutrino experiments,
it is shown that
only two regions of possible values of
$ \left| U_{e3} \right|^2 $
and
$ \left| U_{\mu3} \right|^2 $
are allowed:
I.
$ \left| U_{e3} \right|^2 $
and
$ \left| U_{\mu3} \right|^2 $
are both small and
II.
$ \left| U_{e3} \right|^2 $
is small
and
$ \left| U_{\mu3} \right|^2 $
is large.
If the mixing parameters are in the region I,
$ \nu_\mu \leftrightarrows \nu_e $
oscillations are suppressed.
In this case
the LSND indication in favor of
$ \nu_\mu \leftrightarrows \nu_e $
oscillations is not compatible
with the negative results of all other experiments.
If the mixing parameters are in the region II,
$ \nu_e \leftrightarrows \nu_\tau $
oscillations are strongly suppressed.
If massive neutrinos are Majorana particles,
our analysis shows that
neutrinoless double-beta decay
could be observed in the experiments
of the next generation
only if the mixing parameters are in the region I.
\end{abstract}


\narrowtext

\section{Introduction}
\label{INTRO}

The problem of neutrino masses and mixing
is the most important
problem in neutrino physics today.
The search for the effects of neutrino masses and mixing
is generally considered as one of the important and fertile ways
to probe new scales in physics.
As is well known,
neutrino masses and mixing
naturally arise in GUT models.
The see-saw mechanism
\cite{seesaw}
of Majorana neutrino mass generation,
that is characteristic of GUT models,
is the only known mechanism
that allows one to explain
in a natural way
the smallness of neutrino masses
with respect to the masses
of all the other fundamental fermions.

At present,
there are several
indications in favor of neutrino masses and mixing.
It is a general opinion
(see, for example, Ref.\cite{Winter})
that
the most convincing indications in favor of
a small neutrino mass difference
($ \Delta m^2 \simeq 10^{-5} \, \mathrm{eV}^2 $)
and neutrino mixing
come from the results of the solar neutrino experiments.
In all four solar neutrino experiments
(Homestake, Kamiokande, GALLEX and SAGE)
\cite{solarexp}
the observed event rate is significantly lower
than the event rate predicted
by the Standard Solar Model
(see Refs.\cite{bahcall,turk,cdf}).
A phenomenological
analysis of the data
that does not depend on the predictions
of the Standard Solar Model
indicates that the data of the different experiments
are not compatible with each other
if there are no transitions of solar $\nu_e$'s
into other states
\cite{phenomenological}.

In a recent experiment
\cite{LSND},
the LSND collaboration
found a positive indication
in favor of
$ \bar\nu_\mu \to \bar\nu_e $
oscillations
(see, however, also Ref.\cite{hill}).
The LSND collaboration reported the detection
of nine  $ \bar\nu_e p \to e^{+} n $ events
at a distance of about 30 m
from a target in which
neutrinos
are produced in the decays of stopped
$\pi^{+}$'s
and
$\mu^{+}$'s.
The LSND collaboration calculated
\cite{LSND}
a background of
$ 2.1 \pm 0.3 $
events.
As shown in Ref.\cite{LSND},
there is a
region in the plane of
the parameters
$ \Delta m^2 $ and
$ \sin^2 2\theta $
($\theta$ is the mixing angle)
in which the LSND result
is compatible with
the exclusion plots
obtained in other experiments
\cite{BNLE776,KARMEN}
searching for
$ \nu_\mu \to \nu_e $
transitions.

In this paper we present
an analysis
of the results
of the latest disappearance and appearance
terrestrial neutrino oscillation experiments
in the framework of a general scheme
in which flavor neutrino fields are
mixtures of three neutrino fields
with masses that satisfy
a hierarchy relation.
In this scheme the
oscillations among all active neutrinos
($\nu_{\mu}\leftrightarrows\nu_{e}$,
 $\nu_{\mu}\leftrightarrows\nu_{\tau}$,
 $\nu_{e}\leftrightarrows\nu_{\tau}$)
are rather strongly constrained.
We will show that,
in the region of the values of the mixing parameters
that is allowed by the results of
reactor and accelerator disappearance experiments,
the positive LSND result is not compatible
with the negative results of all
other experiments if there is a hierarchy of couplings
in the lepton sector.

From the results of LEP experiments
which measured the invisible width of the $Z$ boson
(see Ref.\cite{RPP})
it follows that
only three flavor neutrinos exist in nature.
However,
LEP data do not provide
a constraint on the
number of massive light neutrinos.
The number of massive neutrinos
depends on the neutrino mixing scheme.
If the total lepton number is conserved,
neutrinos are Dirac particles
and the number of massive neutrinos is
equal to the number of neutrino flavors.
If the total lepton number is not conserved
and only left-handed neutrino fields enter
in the total Lagrangian
(Majorana mass term),
neutrinos with definite mass are Majorana particles
and the number of massive neutrinos
is again
equal to the number of neutrino flavors.
In the most general case
of neutrino mixing,
left-handed and right-handed
flavor neutrino fields enter in the mass term and
the total lepton number is not conserved
(Dirac and Majorana mass term).
In this case massive
neutrinos are Majorana particles
and
the number of particles
with definite mass is twice
the number of neutrino flavors.
In the case of Dirac and Majorana mass term
there is a very attractive
mechanism which
explains the smallness of the masses of light neutrinos,
i.e.
the see-saw mechanism
\cite{seesaw}.
If one assumes that the total lepton number
is violated at the GUT scale
$M_{\mathrm{GUT}}$
by a right-handed Majorana mass term
and the parameters that characterize the Dirac mass term
are of the order of the corresponding lepton or quark masses,
then in the spectrum of Majorana particles
there are
three very heavy Majorana particles
with masses
$ M_i \simeq M_{\mathrm{GUT}} $
and
three light neutrinos
with masses
$
m_k
\simeq
m_{\mathrm{F}k}^2
/
M_{\mathrm{GUT}}
$
(here $k=1,2,3$
and
$ m_{\mathrm{F}k} $
is the mass of the up-quark or charged lepton
in the $k^{\mathrm{th}}$ generation).
In the see-saw scheme it is expected
that the fields of flavor neutrinos
are predominantly mixtures of
the fields of the three light
Majorana neutrinos.
Thus,
from a general theory of neutrino mixing
it follows that
the possibility of three massive neutrinos
corresponding to the three existing
flavor neutrinos
is quite plausible.

It is to be noted that,
if the number of massive neutrinos is more than three,
oscillations between active and sterile neutrinos
(neutrinos that do not take part in the
standard weak interactions)
are possible
(see, for example, Ref.\cite{Bilenky,CWKim}).
Future solar neutrino experiments
(SNO~\cite{SNO},
Super-Kamiokande~\cite{SK})
could allow to check in a model independent way
\cite{BG95}
if solar
$^8\mathrm{B}$ $\nu_e$'s
transform into sterile states.
Some possible scenarios
with mixing of more than three massive neutrinos
were considered in Ref.\cite{sterile}.

In this paper we will assume that
three flavor neutrino fields
$\nu_{{\alpha}L}$
(with $\alpha=e,\mu,\tau$)
are given by a superposition
of three (Dirac or Majorana)
fields
$\nu_{kL}$
with mass $m_k$ as
\begin{equation}
\nu_{\alpha L}
=
\sum_{k=1}^{3}
U_{\alpha k}
\nu_{kL}
\;.
\label{100}
\end{equation}
Here $U$ is a unitary mixing matrix.

The masses of all the known
fundamental fermions
(leptons and quarks)
satisfy hierarchy relations.
It is natural to assume
that neutrino masses also satisfy
the following hierarchy
\begin{equation}
m_1 \ll m_2 \ll m_3
\;.
\label{103}
\end{equation}
If  neutrino masses are generated with the see-saw mechanism,
the hierarchy relation (\ref{103})
is a consequence of the hierarchy of the
lepton or quark masses.

We will consider
oscillations of terrestrial neutrinos
under the assumption that
the squared mass difference
$ \Delta m^2_{21} \equiv m_2^2 - m_1^2 $
is small
and can be relevant for an explanation
of the deficit of solar $\nu_e$'s
observed by all solar neutrino experiments
\cite{solarexp}
(say
$ \Delta m^2_{21} \simeq 10^{-5} \, \mathrm{eV}^2 $
as suggested by the MSW explanation of the solar neutrino problem
\cite{MSW,SOLMSW}).
Thus,
for all experiments with terrestrial neutrinos we have
$ \Delta m^2_{21} L / 2 p \ll 1 $,
where
$L$ is the distance between
the neutrino source and detector,
and
$p$ is the neutrino momentum.

Under these assumptions
the amplitude of
$ \nu_{\alpha} \to \nu_{\beta} $
transitions
can be written in the following form
\begin{equation}
\cal{A}_{\nu_{\alpha}\to\nu_{\beta}}
=
{\mbox{e}^{ - i E_1 L }}
\left\{
\sum_{k=1,2}
U_{\beta k}
U_{\alpha k}^{*}
+
U_{\beta 3}
U_{\alpha 3}^{*}
\exp
\left(
- i
{\displaystyle
\Delta m^2 L
\over\displaystyle
2 p
}
\right)
\right\}
\;,
\label{201}
\end{equation}
with
$ \Delta m^2 \equiv m^2_3 - m^2_1 $
and
$ \displaystyle
E_k
=
\sqrt{ p^2 + m_k^2 }
\simeq
p
+
{\displaystyle
m_k^2
\over\displaystyle
2 p
}
$.
Then,
taking into account the fact that,
due to the unitarity of the mixing matrix,
\begin{equation}
\sum_{k=1,2}
U_{\beta k}
U_{\alpha k}^{*}
=
\delta_{\alpha\beta}
-
U_{\beta 3}
U_{\alpha 3}^{*}
\;,
\label{202}
\end{equation}
for the probability of
$ \nu_{\alpha} \to \nu_{\beta} $
transitions
with $\beta\not=\alpha$
we obtain
the following expression
(see, for example, Ref.\cite{BFP92})
\begin{equation}
P_{\nu_{\alpha}\to\nu_{\beta}}
=
{1\over2}
\,
A_{\nu_{\alpha};\nu_{\beta}}
\left(
1
-
\cos
{\displaystyle
\Delta m^2 \, L
\over\displaystyle
2 \, p
}
\right)
\;,
\label{107}
\end{equation}
where
\begin{equation}
A_{\nu_{\alpha};\nu_{\beta}}
=
A_{\nu_{\beta};\nu_{\alpha}}
=
4
\left| U_{\alpha3} \right|^2
\left| U_{\beta3} \right|^2
\label{108}
\end{equation}
is the
amplitude of
$ \nu_{\alpha} \leftrightarrows \nu_{\beta} $
oscillations.

The expression for the probability of
$\nu_{\alpha}$
to survive
can be derived from Eq.(\ref{107})
and the conservation of the total probability as
\begin{equation}
P_{\nu_{\alpha}\to\nu_{\alpha}}
=
1
-
\sum_{\beta\not=\alpha}
P_{\nu_{\alpha}\to\nu_{\beta}}
=
1
-
{1\over2}
\,
B_{\nu_{\alpha};\nu_{\alpha}}
\left(
1
-
\cos
{\displaystyle
\Delta m^2 \, L
\over\displaystyle
2 \, p
}
\right)
\;,
\label{109}
\end{equation}
where
the oscillation amplitude
$ B_{\nu_{\alpha};\nu_{\alpha}} $
is given by
\begin{equation}
B_{\nu_{\alpha};\nu_{\alpha}}
=
\sum_{\beta\not=\alpha}
A_{\nu_{\alpha};\nu_{\beta}}
\;.
\label{112}
\end{equation}
Using the unitarity of the mixing matrix,
from Eqs.(\ref{108}) and (\ref{112})
we obtain
\begin{equation}
B_{\nu_{\alpha};\nu_{\alpha}}
=
4
\left| U_{\alpha3} \right|^2
\left(
1
-
\left| U_{\alpha3} \right|^2
\right)
\;.
\label{110}
\end{equation}

We wish to emphasize the following important features of
terrestrial neutrino oscillations
in the model under consideration.

\begin{enumerate}

\item
All oscillation channels
($\nu_{\mu}\leftrightarrows\nu_{e}$,
 $\nu_{\mu}\leftrightarrows\nu_{\tau}$,
 $\nu_{e}\leftrightarrows\nu_{\tau}$)
are open
and the transition probabilities
are determined by three parameters,
$ \Delta m^2 $,
$ \left| U_{e3} \right|^2 $
and
$ \left| U_{\mu3} \right|^2 $
(from the unitarity of the mixing matrix it follows that
$
\left| U_{\tau3} \right|^2
=
1
-
\left| U_{e3} \right|^2
-
\left| U_{\mu3} \right|^2
$).

\item
The oscillations in all channels
are characterized
by the {\em same} oscillation length
$
L_{\mathrm{osc}}
=
4 \pi p / \Delta m^2
$
and the amplitudes
of the inclusive transitions
$ \nu_{\alpha} \to \nu_{\alpha} $
and exclusive transitions
$ \nu_{\alpha} \to \nu_{\beta} $
($\alpha\not=\beta$)
are related by the relation (\ref{112}).

\item
The probabilities of
$ \nu_{\alpha} \to \nu_{\beta} $
and
$ \bar\nu_{\alpha} \to \bar\nu_{\beta} $
are equal,
i.e.
\begin{equation}
P_{\nu_{\alpha}\to\nu_{\beta}}
=
P_{\bar\nu_{\alpha}\to\bar\nu_{\beta}}
\;.
\label{801}
\end{equation}
This relation is a consequence
of the fact that the oscillation probabilities
depend only on the squared moduli
of the elements of the mixing matrix.
In a general case of mixing of three massive neutrinos,
Eq.(\ref{801})
for $\alpha\not=\beta$
is satisfied only if CP is conserved in the lepton sector.
Notice that Eq.(\ref{801})
for $\alpha=\beta$
is a consequence of CPT invariance.

\end{enumerate}

The expressions (\ref{107}) and (\ref{109})
have the same form
as the standard expressions
for the transition probabilities
in the case of mixing between
two massive neutrino fields.
In the latter case
only oscillations between two flavors are possible
and
the oscillation probabilities
are characterized by two parameters,
$ \Delta m^2 $
and
$ \sin^2 2\theta $.
The important difference between
the two-neutrino mixing scheme
and the scheme with
three-neutrino mixing
and a mass hierarchy that we consider
in the present paper
is that
the second scheme
allows simultaneous transitions
among all three flavor neutrinos.
Some preliminary results of the analysis 
of the terrestrial neutrino oscillation experiments
in the framework of this scheme
were already presented in Ref.\cite{BBGK95}
and in this paper we wish to
elaborate the details and present further new results.

\section{Analysis
of the results of neutrino oscillation experiments}
\label{IMPLI}

We will present here an analysis of
the results of the
latest oscillation experiments
with terrestrial neutrinos
in the framework of the model with mixing of three massive neutrinos
and a neutrino mass hierarchy.
We will also consider the results of the LSND experiment,
in which positive indications
in favor of
$ \bar\nu_\mu \leftrightarrows \bar\nu_e $
oscillations were found \cite{LSND}.
The new experiments
CHORUS
\cite{CHORUS}
and NOMAD
\cite{NOMAD}
searching for
$ \nu_\mu \to \nu_\tau $
transitions
are under way at CERN.
These experiments
are sensitive to
$ \Delta m^2 $
in a few eV$^2$ range
(which is of great interest for the problem
of  dark matter in the universe)
and to small values of the oscillation amplitude
($ A_{\nu_{\mu};\nu_{\tau}} \gtrsim 10^{-4} $).
We will discuss
possible implications
of the results of these experiments for
$ \nu_\mu \leftrightarrows \nu_e $
oscillations
after the projected sensitivity will be reached.
We will also consider possible implications
of the results of neutrino oscillation experiments
for neutrinoless $\beta\beta$ decay
($(\beta\beta)_{0\nu}$)
in the case of
massive Majorana neutrinos.

\subsection{Reactor and accelerator disappearance experiments}
\label{DISAPPEARANCE}

Let us start with
the analysis of the results of
reactor and accelerator disappearance experiments
searching for
$ \nu_e \to \nu_{x} $
and
$ \nu_\mu \to \nu_{x} $
transitions.
No indication in favor of
neutrino oscillations was found in these experiments.
We will use the exclusion plots
obtained
in the Bugey reactor experiment
\cite{BUGEY95}
and in the
CDHS and CCFR84 accelerator experiments
\cite{CDHS84,CCFR84}.
At fixed values of
$ \Delta m^2 $,
the allowed values of
the amplitudes
$ B_{\nu_{e};\nu_{e}} $
and
$ B_{\nu_{\mu};\nu_{\mu}} $
are constrained by
\begin{equation}
B_{\nu_{\alpha};\nu_{\alpha}}
\le
B_{\nu_{\alpha};\nu_{\alpha}}^{0}
\qquad
(\alpha=e,\mu)
\;.
\label{141}
\end{equation}
The values of
$ B_{\nu_{e};\nu_{e}}^{0} $
and
$ B_{\nu_{\mu};\nu_{\mu}}^{0} $
can be obtained from the corresponding exclusion curves.

The parameters
$ \left| U_{\alpha3} \right|^2 $
(with $\alpha=e,\mu$)
are expressed in terms of the amplitudes
$ B_{\nu_{\alpha};\nu_{\alpha}} $
as
\begin{equation}
\left| U_{\alpha3} \right|^2
=
{1\over2}
\left(
1
\pm
\sqrt{ 1 - B_{\nu_{\alpha};\nu_{\alpha}} }
\right)
\;.
\label{143}
\end{equation}
It is obvious that,
due to the symmetry of
Eq.(\ref{110})
with respect to the interchange
$ \left| U_{\alpha3} \right|^2 \leftrightarrows
1 - \left| U_{\alpha3} \right|^2 $, 
two values of
$ \left| U_{\alpha3} \right|^2 $
correspond to the same value of
$ B_{\nu_{\alpha};\nu_{\alpha}} $.
Thus,
from the negative results
of reactor and accelerator disappearance experiments,
one can infer that the parameters
$ \left| U_{\alpha3} \right|^2 $
at fixed values of
$ \Delta m^2 $
must satisfy one of the following inequalities
\arraycolsep=3pt
\begin{eqnarray}
&&
\left| U_{\alpha3} \right|^2
\le
a^{0}_{\alpha}
\label{802}
\\
\mathrm{or}
&&
\nonumber
\\
&&
\left| U_{\alpha3} \right|^2
\ge
1 - a^{0}_{\alpha}
\;,
\label{803}
\end{eqnarray}
with
\begin{equation}
a^{0}_{\alpha}
\equiv
{1\over2}
\left(
1
-
\sqrt{ 1 - B_{\nu_{\alpha};\nu_{\alpha}}^{0} }
\right)
\;.
\label{804}
\end{equation}

We will consider values of
$ \Delta m^2 $
in the interval,
$ 10^{-1} \, \mathrm{eV}^2
\le \Delta m^2 \le
10^{3} \, \mathrm{eV}^2 $,
that covers the range
where positive indications
in favor of
$ \nu_\mu \leftrightarrows \nu_e $
oscillations
were reported by the LSND collaboration
\cite{LSND}.
The values of the parameters
$ a^{0}_{e} $
and
$ a^{0}_{\mu} $
in this interval of
$ \Delta m^2 $,
that were obtained from
the exclusion plots of
the Bugey experiment \cite{BUGEY95}
and of the CDHS and CCFR84 experiments \cite{CDHS84,CCFR84},
respectively,
are presented
in Fig.\ref{aelamu}.
It can be seen from Fig.\ref{aelamu}
that in the region of
$ \Delta m^2 $
under consideration
$ a^{0}_{e} $
is always small
and
$ a^{0}_{\mu} $ is small for
$ \Delta m^2 \gtrsim 0.5 \, \mathrm{eV}^2 $.

From Eqs.(\ref{802}) and (\ref{803})
it follows that
the parameters
$ \left| U_{\alpha3} \right|^2 $
are either small or large
(i.e. close to one).
However,
due to the unitarity of the mixing matrix
the parameters
$ \left| U_{e3} \right|^2 $
and
$ \left| U_{\mu3} \right|^2 $
must satisfy the inequality
$ \left| U_{e3} \right|^2 
+
\left| U_{\mu3} \right|^2 \le 1 $
and
consequently they
cannot be both large.
Thus,
on the basis of an analysis of the results of
only the reactor and accelerator
disappearance experiments,
we reach the conclusion that
the values of the parameters
$ \left| U_{e3} \right|^2 $
and
$ \left| U_{\mu3} \right|^2 $
can lie in one of the following three regions:

\def\theenumi{\Roman{enumi}}

\begin{enumerate}

\item \label{R1}
The region of small
$ \left| U_{e3} \right|^2 $
and
$ \left| U_{\mu3} \right|^2 $.

\item \label{R2}
The region of small
$ \left| U_{e3} \right|^2 $
and large
$ \left| U_{\mu3} \right|^2 $.

\item \label{R3}
The region of large
$ \left| U_{e3} \right|^2 $
and small
$ \left| U_{\mu3} \right|^2 $.

\end{enumerate}

It can be shown \cite{BBGK95}
that
the region \ref{R3} is excluded by the
results of solar neutrino experiments.
In fact,
in the model under consideration
the survival probability of the solar $\nu_e$'s
is given by
\cite{SOLARTHREEGEN}
\begin{equation}
P_{\nu_e\to\nu_e}
=
\left(
1
-
\left| U_{e3} \right|^2
\right)^2
P_{\nu_e\to\nu_e}^{(1,2)}
+
\left| U_{e3} \right|^4
\;,
\label{172}
\end{equation}
where
$ P_{\nu_e\to\nu_e}^{(1,2)} $
is the survival probability
due to the mixing between
the first and the second generations.
If the parameter
$ \left| U_{e3} \right|^2 $
is large,
from Eq.(\ref{172})
in the region of
$ \Delta m^2 $
under consideration
we have
$
P_{\nu_e\to\nu_e}
\ge
0.92
$
for all values of the neutrino energy.
This large lower bound of the $\nu_e$ survival probability
is not compatible
with the results of solar neutrino experiments
\cite{solarexp}.

In the following, we will consider, in detail, the
region \ref{R1} and the region \ref{R2}.
We will take into account
the results of all the latest appearance experiments
searching for neutrino oscillations.

\subsection{The region of small
$ \left| U_{e3} \right|^2 $
and
$ \left| U_{\mu3} \right|^2 $}

In the region \ref{R1},
the parameters
$ \left| U_{e3} \right|^2 $
and
$ \left| U_{\mu3} \right|^2 $
are small and bounded as
\begin{equation}
\left| U_{e3} \right|^2
\le
a^{0}_{e}
\quad
\mathrm{and}
\quad
\left| U_{\mu3} \right|^2
\le
a^{0}_{\mu}
\;,
\label{600}
\end{equation}
where the quantities
$ a^{0}_{e} $ and $ a^{0}_{\mu} $
are given by Eq.(\ref{804})
and
in the interval of
$ \Delta m^2 $
under consideration,
take the values as presented
in Fig.\ref{aelamu}.
The region \ref{R1} is of great theoretical interest.
It is now  well established that
the non-diagonal elements of the CKM mixing matrix
of quarks, $V$, 
are small and satisfy the  hierarchy
$
\left| V_{13} \right|
\ll
\left| V_{23} \right|
\ll
\left| V_{12} \right|
$.
Is this feature
common to the quark and lepton sectors?
A hierarchy of couplings in the lepton sector
similar to the one in the quark sector
can be realized
only if the values of the parameters
$ \left| U_{e3} \right|^2 $
and
$ \left| U_{\mu3} \right|^2 $
lie in the region \ref{R1}.

If the values of the parameters
$ \left| U_{e3} \right|^2 $
and
$ \left| U_{\mu3} \right|^2 $
are in the region \ref{R1},
we can expect that
$ \nu_\mu \leftrightarrows \nu_e $
oscillations are suppressed
with respect to
$ \nu_\mu \leftrightarrows \nu_\tau $
and
$ \nu_e \leftrightarrows \nu_\tau $
oscillations.
In fact,
in the region \ref{R1},
the amplitude of
$ \nu_\mu \leftrightarrows \nu_e $
oscillations is given by a product of  two small quantities
$ \left| U_{e3} \right|^2 $
and
$ \left| U_{\mu3} \right|^2 $,
whereas the amplitudes of
$ \nu_\mu \leftrightarrows \nu_\tau $
and
$ \nu_e \leftrightarrows \nu_\tau $
oscillations
are linear in one of these
two small quantities
(see Eq.(\ref{108})).

Let us consider, in detail, 
$ \nu_\mu \leftrightarrows \nu_e $
oscillations.
From the results of
reactor and accelerator disappearance experiments,
we obtain
the following upper bound
for the oscillation amplitude
\begin{equation}
A_{\nu_{\mu};\nu_{e}}
\le
4 \, a^{0}_{e} \, a^{0}_{\mu}
\;.
\label{601}
\end{equation}
In Fig.\ref{r1amuel},
we have plotted the curve
that represents this
upper bound,
which was obtained
from the results
of the Bugey \cite{BUGEY95},
CDHS \cite{CDHS84} and CCFR84 \cite{CCFR84} experiments
(the curve passing through the circles).
In Fig.\ref{r1amuel}
we have also plotted the exclusion curves
obtained in the
BNL E776
\cite{BNLE776}
(dash-dotted line)
and
KARMEN
\cite{KARMEN}
(dash-dot-dotted line)
experiments
searching for 
$ \nu_\mu \to \nu_e $
transitions.
The region allowed by the results of the LSND experiment
is shown in Fig.\ref{r1amuel}
as the shadowed region between the two solid lines.
Taking into account that
$
A_{\nu_{\mu};\nu_{e}}
\le
B_{\nu_{e};\nu_{e}}
$
(see Eq.(\ref{112})),
we have also plotted, in Fig.\ref{r1amuel}, 
the exclusion curve
for
$ B_{\nu_{e};\nu_{e}} $
found in the Bugey experiment
(dashed line).
It can be seen from the figure that
for small values of
$ \Delta m^2 $
($ \Delta m^2 \lesssim 0.5 \, \mathrm{eV}^2 $)
this bound on
$ A_{\nu_{\mu};\nu_{e}} $
is stronger than the direct bound
obtained by the BNL E776 and KARMEN experiments.
It is also seen from the figure that
this bound is not compatible with the result
of the LSND experiment
for
$ \Delta m^2 \lesssim 0.2 \, \mathrm{eV}^2 $.

Fig.\ref{r1amuel} shows 
that,
in the range of
$ \Delta m^2 $
under consideration,
with the exception of the interval
$ 10 \, \mathrm{eV}^2
\lesssim \Delta m^2 \lesssim
60 \, \mathrm{eV}^2 $,
the limits on the oscillation amplitude
$ A_{\nu_{\mu};\nu_{e}} $
that can be obtained
from the results of disappearance experiments
are more stringent than the limits
obtained from the experiments searching for
$ \nu_\mu \to \nu_e $
transitions.
From Fig.\ref{r1amuel},
it can be seen that
the LSND result
is not compatible
with the results of the reactor and accelerator
disappearance experiments
in the range of
$ \Delta m^2 $
under consideration,
with the exception of the interval
$ 5 \, \mathrm{eV}^2
\lesssim \Delta m^2 \lesssim
70 \, \mathrm{eV}^2 $.
In this interval of
$ \Delta m^2 $
the result of the LSND experiment
is only marginally compatible
with the exclusion curve
obtained in the BNL E776 experiment.

Thus,
from the results of disappearance experiments
we have obtained rather strong limits
on the allowed values of the oscillation amplitude
$ A_{\nu_{\mu};\nu_{e}} $
in the region \ref{R1}
of the parameters
$ \left| U_{e3} \right|^2 $
and
$ \left| U_{\mu3} \right|^2 $.
We will take now into account
also the results of the FNAL E531 experiment
\cite{FNALE531}
on the search for
$ \nu_\mu \to \nu_\tau $
and
$ \nu_e \to \nu_\tau $
transitions
and
the recent results of the CCFR95 experiment
\cite{CCFR95}
on the search for
$ \nu_\mu \to \nu_\tau $
transitions.
From the exclusion plots deduced from
these two experiments,
for a fixed value of
$ \Delta m^2 $,
the values of
$ A_{\nu_{\mu};\nu_{\tau}} $
and
$ A_{\nu_{e};\nu_{\tau}} $
are constrained by
\arraycolsep=0pt
\begin{eqnarray}
&&
A_{\nu_{\mu};\nu_{\tau}}
\le
A_{\nu_{\mu};\nu_{\tau}}^{0}
\;,
\label{303}
\\
&&
A_{\nu_{e};\nu_{\tau}}
\le
A_{\nu_{e};\nu_{\tau}}^{0}
\;.
\label{304}
\end{eqnarray}

From Eq.(\ref{108})
it follows that
in the linear approximation in the small quantities
$ \left| U_{e3} \right|^2 $
and
$ \left| U_{\mu3} \right|^2 $
we have
\arraycolsep=0pt
\begin{eqnarray}
&&
\left| U_{\mu3} \right|^2
\simeq
{1\over4}
\,
A_{\nu_{\mu};\nu_{\tau}}
\;,
\label{307}
\\
&&
\left| U_{e3} \right|^2
\simeq
{1\over4}
\,
A_{\nu_{e};\nu_{\tau}}
\;.
\label{308}
\end{eqnarray}
Thus,
the parameters
$ \left| U_{\mu3} \right|^2 $
and
$ \left| U_{e3} \right|^2 $
are determined, respectively, by the amplitudes of
$ \nu_{\mu} \leftrightarrows \nu_{\tau} $
and
$ \nu_{e} \leftrightarrows \nu_{\tau} $
oscillations.
With the help of Eq.(\ref{307}),
from
the results of the FNAL E531 and CCFR95 experiments
it is possible to obtain,
for
$ \Delta m^2 \gtrsim 4 \, \mathrm{eV}^2 $,
more stringent limits
on the value of
$ \left| U_{\mu3} \right|^2 $
than those obtained
from the results of the
CDHS and CCFR84 disappearance experiments.
Combining these limits
with the limits on
$ \left| U_{e3} \right|^2 $
obtained from the results of the Bugey
disappearance experiment,
we have found rather stringent
limits on the value of the amplitude
$ A_{\nu_{\mu};\nu_{e}} $.
In fact,
from Eqs.(\ref{600}) and (\ref{307})
we have
\begin{equation}
A_{\nu_{\mu};\nu_{e}}
\lesssim
a^{0}_{e}
\,
A_{\nu_{\mu};\nu_{\tau}}^{0}
\;.
\label{725}
\end{equation}
The bound (\ref{725}) on the amplitude
$ A_{\nu_{\mu};\nu_{e}} $
obtained
from the results of the Bugey, FNAL E531 and CCFR95 experiments
is presented in Fig.\ref{r1amuel}
(the curve passing through the triangles).
It can be seen in Fig.\ref{r1amuel} that
the results of reactor and accelerator
disappearance experiments,
together with the results of
$ \nu_\mu \to \nu_\tau $
appearance experiments
exclude
all the region of the parameters
$ \Delta m^2 $
and
$ A_{\nu_{\mu};\nu_{e}} $
that is allowed by the LSND experiment.
It is to be emphasized that this result has been derived
under the assumption that
the parameters
$ \left| U_{e3} \right|^2 $
and
$ \left| U_{\mu3} \right|^2 $
are both small.

In Fig.\ref{r1amuel},
we have also shown the region in the
$ A_{\nu_{\mu};\nu_{e}} $--$ \Delta m^2 $
plane that could be explored
when the projected sensitivity of the
CHORUS \cite{CHORUS} and NOMAD \cite{NOMAD}
experiments,
which are searching for
$ \nu_\mu \to \nu_\tau $
transitions,
is reached
(the region delimited by the line passing through the squares).

From Eqs.(\ref{307}) and (\ref{308})
we obtain the following inequality
\begin{equation}
A_{\nu_{\mu};\nu_{e}}
\lesssim
{1\over4}
\,
A_{\nu_{\mu};\nu_{\tau}}^{0}
\,
A_{\nu_{e};\nu_{\tau}}^{0}
\;.
\label{825}
\end{equation}
The limits obtained,
by using this inequality,
from the results of the FNAL E531 and CCFR95 experiments
on the search for
$ \nu_\mu \to \nu_\tau $
transitions
and
from the results of the FNAL E531 experiment on
the search for
$ \nu_e \to \nu_\tau $
transitions
are presented in Fig.\ref{r1amuel}
(the dotted line).

Finally,
since in the region \ref{R1}
the amplitude
$ A_{\nu_{\mu};\nu_{e}} $
is very small,
we obtain, from Eq.(\ref{112}),
the following relations
between the amplitudes
of inclusive and exclusive transitions
\arraycolsep=0pt
\begin{eqnarray}
&&
B_{\nu_{\mu};\nu_{\mu}}
\simeq
A_{\nu_{\mu};\nu_{\tau}}
\;,
\label{310}
\\
&&
B_{\nu_{e};\nu_{e}}
\simeq
A_{\nu_{e};\nu_{\tau}}
\;.
\label{311}
\end{eqnarray}
A test of these relations
will allow us to check
if there is a hierarchy
of masses and couplings in the lepton sector.

\subsection{The region of small
$ \left| U_{e3} \right|^2 $
and large
$ \left| U_{\mu3} \right|^2 $}

We will now consider in detail
the region \ref{R2}
in which the parameter
$ \left| U_{e3} \right|^2 $
is small and
$ \left| U_{\mu3} \right|^2 $
is large (close to one),
i.e.
\begin{equation}
\left| U_{e3} \right|^2
\le
a_{e}^{0}
\quad \mathrm{and} \quad
\left| U_{\mu3} \right|^2
\ge
1 - a_{\mu}^{0}
\;,
\label{400}
\end{equation}
with the quantities
$a_{e}^{0}$ and $a_{\mu}^{0}$
given by Eq.(\ref{804}).

In this region,
$\nu_{e}\leftrightarrows\nu_{\tau}$
oscillations
are strongly suppressed with respect to
$\nu_{\mu}\leftrightarrows\nu_{e}$
and
$\nu_{\mu}\leftrightarrows\nu_{\tau}$
oscillations.
This is due to the fact that
the amplitude
$ A_{\nu_{e};\nu_{\tau}} $
is quadratic in the small quantities
$ \left| U_{e3} \right|^2 $
and
$ ( 1 - \left| U_{\mu3} \right|^2 ) $,
whereas the amplitudes
$ A_{\nu_{\mu};\nu_{e}} $
and
$ A_{\nu_{\mu};\nu_{\tau}} $
are linear
(see Eq.(\ref{108})).

Taking into account the unitarity bound
$ \left| U_{e3} \right|^2 \le
1 - \left| U_{\mu3} \right|^2 $,
it follows from Eq.(\ref{400}) that
the parameter
$ \left| U_{e3} \right|^2 $
must also satisfy  the following inequality
\begin{equation}
\left| U_{e3} \right|^2
\le
a_{\mu}^{0}
\;.
\label{444}
\end{equation}
From Eqs.(\ref{400}) and (\ref{444}),
we obtain the following upper bound
for the amplitude of
$ \nu_e \leftrightarrows \nu_\tau $
oscillations
\begin{equation}
A_{\nu_{e};\nu_{\tau}}
\le
4
\,
\mathrm{Min}\left[ a_{e}^{0} , a_{\mu}^{0} \right]
\,
a_{\mu}^{0}
\;.
\label{404}
\end{equation}
From this inequality
it follows that
$ \nu_e \leftrightarrows \nu_\tau $
oscillations are strongly suppressed
in the region \ref{R2}.
In Fig.\ref{r2aelta},
we have plotted the upper bound
for the amplitude
$ A_{\nu_{e};\nu_{\tau}} $
obtained
from Eq.(\ref{404}) using the results
of the Bugey, CDHS and CCFR84 experiments
(the curve passing through the triangles).
From this figure it can be seen that
in all the considered range of $ \Delta m^2 $
the upper bound for $ A_{\nu_{e};\nu_{\tau}} $
is very small,
varying from about $10^{-4}$ to about $4\times10^{-2}$.

Up to now we have used the limits
on the parameters
$ \left| U_{e3} \right|^2 $
and
$ \left| U_{\mu3} \right|^2 $
that were obtained
from the results of disappearance experiments.
We will now take into account also the
results of the BNL E776 experiment
\cite{BNLE776} that has searched for
$ \nu_\mu \to \nu_e $
transitions.
From the exclusion plot obtained
in this experiment,
it follows that,
for a fixed value of
$ \Delta m^2 $,
the value of the
$ \nu_\mu \leftrightarrows \nu_e $
oscillation amplitude
is constrained by
\begin{equation}
A_{\nu_{\mu};\nu_{e}}
\le
A_{\nu_{\mu};\nu_{e}}^{0}
\;.
\label{375}
\end{equation}

Since in the region \ref{R2}
under consideration,
the parameter
$ \left| U_{\mu3} \right|^2 $
is close to one
in the linear approximation in the small quantities
$ \left| U_{e3} \right|^2 $
and
$ \left( 1 - \left| U_{\mu3} \right|^2 \right) $,
we have
\begin{equation}
A_{\nu_{\mu};\nu_{e}}
\simeq
4 \, \left| U_{e3} \right|^2
\;.
\label{405}
\end{equation}
From this relation and Eq.(\ref{375}) we have
\begin{equation}
\left| U_{e3} \right|^2
\lesssim
{\displaystyle
A_{\nu_{\mu};\nu_{e}}^{0}
\over\displaystyle
4
}
\;.
\label{410}
\end{equation}
The upper bound for
$ \left| U_{e3} \right|^2 $
obtained from the results
of the BNL E776 experiment
are presented in Fig.\ref{r2uel}
(the dashed curve).
From this figure it can be seen that
for
$ 0.7 \, \mathrm{eV}^2
\lesssim \Delta m^2 \lesssim
10^{3} \, \mathrm{eV}^2 $
the limits on the parameter
$ \left| U_{e3} \right|^2 $
that can be found from the results of the BNL E776 experiment
are much more stringent than those obtained from the
exclusion plot of the Bugey experiment
(dash-dotted curve)
and from the results of the CDHS and CCFR experiments
with Eq.(\ref{444})
(dash-dot-dotted curve).

From the positive results of the LSND experiment,
we can find a region of {\em allowed} values of
$ \left| U_{e3} \right|^2 $
in the
$ \left| U_{e3} \right|^2 $--$ \Delta m^2 $
plane.
In fact,
for a fixed value of $ \Delta m^2 $,
from the LSND allowed region in the
$ A_{\nu_{\mu};\nu_{e}} $--$ \Delta m^2 $
plane,
we have
\begin{equation}
A_{\nu_{\mu};\nu_{e}}^{(-)}
\le
A_{\nu_{\mu};\nu_{e}}
\le
A_{\nu_{\mu};\nu_{e}}^{(+)}
\;.
\label{302}
\end{equation}
Hence, with Eq.(\ref{405})
we obtain
\begin{equation}
{\displaystyle
A_{\nu_{\mu};\nu_{e}}^{(-)}
\over\displaystyle
4
}
\lesssim
\left| U_{e3} \right|^2
\lesssim
{\displaystyle
A_{\nu_{\mu};\nu_{e}}^{(+)}
\over\displaystyle
4
}
\;.
\label{411}
\end{equation}
The corresponding allowed region
is presented in Fig.\ref{r2uel}
(the shadowed region between the solid lines).

From the inequality (\ref{410}),
it is possible to derive
the following upper bound
for the amplitude of
$ \nu_e \leftrightarrows \nu_\tau $
oscillations
\begin{equation}
A_{\nu_{e};\nu_{\tau}}
\lesssim
A_{\nu_{\mu};\nu_{e}}^{0}
\,
a_{\mu}^{0}
\;.
\label{425}
\end{equation}
In Fig.\ref{r2aelta}
we have plotted
the corresponding boundary curve
obtained from the results
of the CDHS, CCFR84 and BNL E776 experiments
(the curve passing through the squares).
From this figure it can be seen
that for
$ \Delta m^2 \gtrsim 1 \, \mathrm{eV}^2 $
the upper bound on the oscillation amplitude
$ A_{\nu_{e};\nu_{\tau}} $
varies from $ 2 \times 10^{-5} $ to $ 5 \times 10^{-4} $
and
is more stringent than that obtained
from the results of disappearance experiments
using Eq.(\ref{404}).

From the results of the experiments  searching for
$ \nu_\mu \to \nu_\tau $
transitions,
we can obtain even stronger limits
on the amplitude of
$ \nu_e \leftrightarrows \nu_\tau $
oscillations.
In fact,
in the region \ref{R2}
under consideration,
where
$ \left| U_{\mu3} \right|^2 \simeq 1 $,
the negative results of the experiments
searching for
$ \nu_\mu \to \nu_\tau $
transitions
give the following upper bound on the value of
$ \left| U_{\tau3} \right|^2 $
\begin{equation}
\left| U_{\tau3} \right|^2
\lesssim
{\displaystyle
A_{\nu_{\mu};\nu_{\tau}}^{0}
\over\displaystyle
4
}
\;.
\label{460}
\end{equation}
From Eqs.(\ref{410}) and (\ref{460})
we have the following constraint on
the amplitude of
$ \nu_e \leftrightarrows \nu_\tau $
oscillations
\begin{equation}
A_{\nu_{e};\nu_{\tau}}
\lesssim
{\displaystyle
A_{\nu_{\mu};\nu_{e}}^{0}
\,
A_{\nu_{\mu};\nu_{\tau}}^{0}
\over\displaystyle
4
}
\;.
\label{502}
\end{equation}
The corresponding boundary curve obtained
from the results
of the BNL E776, FNAL E531 and CCFR95 experiments
is presented in Fig.\ref{r2aelta}
(the curve passing through the circles).
From this figure it can be seen that for
$ \Delta m^2 \gtrsim 4 \, \mathrm{eV}^2 $
the amplitude of
$ \nu_e \leftrightarrows \nu_\tau $
oscillations
is extremely small
(less than $ 3 \times 10^{-5} $).

Furthermore, from Eqs.(\ref{410}) and (\ref{460}),
we have
\begin{equation}
1 - \left| U_{\mu3} \right|^2
\lesssim
{\displaystyle
A_{\nu_{\mu};\nu_{e}}^{0}
+
A_{\nu_{\mu};\nu_{\tau}}^{0}
\over\displaystyle
4
}
\;.
\label{501}
\end{equation}
The upper bound for
$ 1 - \left| U_{\mu3} \right|^2 $
obtained with Eq.(\ref{501}) from the results
of the BNL E776, FNAL E531 and CCFR95 experiments
is presented in Fig.\ref{r2umu}
(the dashed line).
The solid line in Fig.\ref{r2umu}
represents the upper bound (\ref{400})
obtained from the results of the CDHS and CCFR84 experiments.
This figure shows that
the inequality (\ref{501}) enables us
to obtain,
in a wide range of values of
$ \Delta m^2 $,
more stringent limits
on the values of the parameter
$ \left| U_{\mu3} \right|^2 $
from the results of the experiments
on the search for
$ \nu_\mu \to \nu_e $
and
$ \nu_\mu \to \nu_\tau $
transitions
than those obtained from the results
of $\nu_\mu$ disappearance experiments.

Finally,
the fact that
the amplitude
$ A_{\nu_{e};\nu_{\tau}} $
is very small leads to
the following relations
between the amplitudes of inclusive
$ \nu_\alpha \to \nu_\alpha $
and exclusive
$ \nu_\alpha \to \nu_\beta $
($ \beta \not= \alpha $)
transitions
\arraycolsep=0pt
\begin{eqnarray}
&&
B_{\nu_{e};\nu_{e}}
\simeq
A_{\nu_{\mu};\nu_{e}}
\;,
\label{551}
\\
&&
B_{\nu_{\mu};\nu_{\mu}}
=
A_{\nu_{\mu};\nu_{e}}
+
A_{\nu_{\mu};\nu_{\tau}}
\;.
\label{555}
\end{eqnarray}
Thus,
in the region \ref{R2}
the amplitudes
$ B_{\nu_{e};\nu_{e}} $
and
$ B_{\nu_{\mu};\nu_{\mu}} $
are determined by
the amplitudes
$ A_{\nu_{\mu};\nu_{e}} $
and
$ A_{\nu_{\mu};\nu_{\tau}} $.
From the results of the LSND experiment and 
using
Eqs.(\ref{303}), (\ref{375}),
(\ref{302}), (\ref{551}) and (\ref{555}),
we obtain
\arraycolsep=0pt
\begin{eqnarray}
&&
A_{\nu_{\mu};\nu_{e}}^{(-)}
\lesssim
B_{\nu_{e};\nu_{e}}
\lesssim
\mathrm{Min}\left[
A_{\nu_{\mu};\nu_{e}}^{(+)}
,
A_{\nu_{\mu};\nu_{e}}^{0}
\right]
\;,
\label{552}
\\
&&
A_{\nu_{\mu};\nu_{e}}^{(-)}
\le
B_{\nu_{\mu};\nu_{\mu}}
\le
\mathrm{Min}\left[
A_{\nu_{\mu};\nu_{e}}^{(+)}
,
A_{\nu_{\mu};\nu_{e}}^{0}
\right]
+
A_{\nu_{\mu};\nu_{\tau}}^{0}
\;.
\label{553}
\end{eqnarray}
A test of these relations
in future reactor and accelerator experiments
could allow us to check the indications
in favor of
$ \nu_\mu \leftrightarrows \nu_e $
oscillations
found by the LSND collaboration.

\section{Neutrinoless $\beta\beta$ decay}

As we have shown in Section \ref{DISAPPEARANCE},
in the framework of the model under consideration,
the results of the solar neutrino experiments
and
those of the reactor and accelerator
neutrino oscillation disappearance experiments
indicate that the parameter
$ \left| U_{e3} \right|^2 $
is small.
If massive neutrinos are Majorana particles,
this fact can have important consequences for
$(\beta\beta)_{0\nu}$ decay experiments.
The matrix element of
$(\beta\beta)_{0\nu}$ decay
is proportional to
$ \displaystyle
\langle m \rangle
=
\sum_{i}
U_{ei}^2
\,
m_{i}
$
(see, for example, Refs.\cite{Bilenky,CWKim}).
The results of the experiments
searching  for
$(\beta\beta)_{0\nu}$ decay
can be summarized as 
$ \left| \langle m \rangle \right| \lesssim 1 \, \mathrm{eV} $
(see Ref.\cite{DOUBLEBETAEXP}).
The expected sensitivity of the next generation
of experiments is
$ \left| \langle m \rangle \right| \simeq 10^{-1} \, \mathrm{eV} $
\cite{DOUBLEBETAEXP}.

In the framework of the model under consideration,
the hierarchy of neutrino masses
implies that
\cite{Petcov94}
\begin{equation}
\left| \langle m \rangle \right|
\simeq
\left| U_{e3} \right|^2
\,
\sqrt{ \Delta m^2 }
\;.
\label{651}
\end{equation}
In Fig.\ref{r1dbeta}
we have plotted the boundary curve
for
$ \left| \langle m \rangle \right| $
obtained with Eq.(\ref{651}) from the results
of the Bugey experiment
in the interval
$ 10^{-1} \, \mathrm{eV}^2
\le \Delta m^2 \le
10^{3} \, \mathrm{eV}^2 $
(the dash-dotted line).
From this figure
it can be seen that
for
$ \Delta m^2 \lesssim 5 \, \mathrm{eV}^2 $,
the upper bound for
$ \left| \langle m \rangle \right| $
is less than the projected sensitivity of the
$(\beta\beta)_{0\nu}$ decay experiments
of the next generation.

In the region \ref{R2},
where
$ \left| U_{e3} \right|^2 $
is small and
$ \left| U_{\mu3} \right|^2 $
is large,
the value of
$ \left| U_{e3} \right|^2 $
is constrained
not only by Eq.(\ref{400}),
but also by Eq.(\ref{444}),
which takes into account the unitarity constraint.
In Fig.\ref{r2dbeta}
we have plotted the corresponding curve
obtained from the results of the CDHS and CCFR84 experiments
(the dash-dot-dotted line).
Furthermore,
in region \ref{R2}
the value of
$ \left| U_{e3} \right|^2 $
is severely constrained by the results of the BNL E776 experiment
that searched for
$ \nu_\mu \to \nu_e $
transitions
(see Eq.(\ref{410})).
The corresponding boundary curve
is presented in Fig.\ref{r2dbeta}
(the dashed curve).
From this figure it can be seen that
$ \left| \langle m \rangle \right| \lesssim 10^{-2} \, \mathrm{eV} $
practically in all the considered range of $ \Delta m^2 $.
Thus,
if the parameters
$ \left| U_{e3} \right|^2 $
and
$ \left| U_{\mu3} \right|^2 $
are in the region \ref{R2},
the observation of
$(\beta\beta)_{0\nu}$ decay becomes a formidable, 
if not impossible,
task.

Finally,
from the results of the LSND experiment,
for each value of $ \Delta m^2 $,
an allowed range of
$ \left| U_{e3} \right|^2 $
can be determined from Eq.(\ref{411})
in region \ref{R2}.
In Fig.\ref{r2dbeta} we have also plotted the corresponding allowed
region in the
$ \Delta m^2 $--$ \left| \langle m \rangle \right| $
plane
(the shadowed region between the two solid lines).

\section{Conclusions}

We have considered a scheme with mixing
of three massive neutrino fields
and a hierarchy of neutrino masses.
We have assumed that
the squared mass difference
$ m_2^2 - m_1^2 $
is small
and can be relevant for the suppression of
solar $\nu_e$'s. 
In this scheme the oscillations of
terrestrial neutrinos in short-baseline experiments
are determined by the values of three parameters,
one squared mass difference
$ \Delta m^2 \equiv m_3^2 - m_1^2 $
and
the squared moduli of
the two mixing matrix elements
$U_{e3}$ and $U_{\mu3}$.

After the calibration of the GALLEX detector
with a radioactive source
\cite{GALLEX}
the indications in favor of neutrino oscillations
coming from solar neutrino experiments
have become more significant. For this reason,
we believe that
the model considered here is
the simplest and most realistic
model of neutrino mixing.
It seems very appropriate to
analyze in the framework of this model
all the data from the existing experiments
that have  searched for neutrino oscillations
and
to infer predictions for the results
of future experiments
(see also Refs.\cite{Lisi,Minakata}).
In our discussion, we have also taken into account
the positive indications
in favor of
$ \nu_\mu \leftrightarrows \nu_e $
oscillations that were found in the recent LSND experiment.

We have shown, by using the results of
reactor and accelerator disappearance experiments,
that
in a wide range of $ \Delta m^2 $
($ 0.5 \, \mathrm{eV}^2
\le \Delta m^2 \le
10^{3} \, \mathrm{eV}^2 $)
the parameters
$ \left| U_{e3} \right|^2 $
and
$ \left| U_{\mu3} \right|^2 $
can be either very small or very large (close to 1).
From the unitarity of the mixing matrix
and
from the results of the solar neutrino experiments
it follows that
only two regions for the parameters
$ \left| U_{e3} \right|^2 $
and
$ \left| U_{\mu3} \right|^2 $
are allowed:

\begin{enumerate}

\item
The region of small
$ \left| U_{e3} \right|^2 $
and
$ \left| U_{\mu3} \right|^2 $.

\item
The region of small
$ \left| U_{e3} \right|^2 $
and large
$ \left| U_{\mu3} \right|^2 $.

\end{enumerate}

The overall situation for the allowed regions in the present
theoretical framework is summarized in
Figs.\ref{uelumu1} and \ref{uelumu2}.
In these figures the
various constraints and the allowed regions
for the values of the parameters
$ \left| U_{e3} \right|^2 $
and
$ \left| U_{\mu3} \right|^2 $
are displayed
at the representative value
$ \Delta m^2 = 6 \, \mbox{eV}^2 $.

If the parameters
$ \left| U_{e3} \right|^2 $
and
$ \left| U_{\mu3} \right|^2 $
are in the region \ref{R1},
$ \nu_\mu \leftrightarrows \nu_e $
oscillations are strongly suppressed.
We have shown that in the region \ref{R1}
the indications
in favor of
$ \nu_\mu \leftrightarrows \nu_e $
oscillations found in the LSND experiment
are not compatible with the results of all other experiments
which have not found any evidence of neutrino oscillations.
It is to be emphasized that
a hierarchy of couplings in the lepton sector
analogous to that of the quark sector
is possible only if the parameters
$ \left| U_{e3} \right|^2 $
and
$ \left| U_{\mu3} \right|^2 $
lie in region \ref{R1}.

If the parameters
$ \left| U_{e3} \right|^2 $
and
$ \left| U_{\mu3} \right|^2 $
are in the region \ref{R2},
the amplitude of
$ \nu_e \leftrightarrows \nu_\tau $
oscillations is very small
in the wide range of $ \Delta m^2 $
under consideration.
In this case there is no constraint on the amplitudes of
$ \nu_\mu \leftrightarrows \nu_e $
and
$ \nu_\mu \leftrightarrows \nu_\tau $
oscillations
and the amplitudes that characterize
the survival probabilities of $\nu_e$'s and $\nu_\mu$'s
are determined by the amplitudes
$A_{\nu_{\mu};\nu_{e}}$
and
$A_{\nu_{\mu};\nu_{\tau}}$.
The indication
in favor of
$ \nu_\mu \leftrightarrows \nu_e $
oscillations found in the LSND experiment
can be checked in future disappearance experiments.

We have also discussed the implications
of our analysis for
$(\beta\beta)_{0\nu}$ decay
if massive neutrinos turn out to be Majorana particles.
If the parameters
$ \left| U_{e3} \right|^2 $
and
$ \left| U_{\mu3} \right|^2 $
lie in the region \ref{R1} and
$ \Delta m^2 \gtrsim 5 \, \mathrm{eV}^2 $,
$(\beta\beta)_{0\nu}$ decay
could be observed in the next generation of experiments.
If the parameters
$ \left| U_{e3} \right|^2 $
and
$ \left| U_{\mu3} \right|^2 $
are in the region \ref{R2},
the sensitivity of the $(\beta\beta)_{0\nu}$ decay experiments
of the next generation
is not sufficient to observe this process.

\acknowledgments

It is a pleasure for us
to express our gratitude
to
Serguey Petcov
for very useful discussions.
S.B. would like to acknowledge
with gratefulness the kind hospitality of the
Elementary Particle Sector of SISSA,
where part of this work has been done.

\begin{figure}[h]
\protect\caption{Values of the parameters
$ a^{0}_{e} $
and
$ a^{0}_{\mu} $
(see Eq.(\protect\ref{804}))
obtained from the results
of reactor and accelerator disappearance experiments
for
$ \Delta m^2 $
in the range
$ 10^{-1} \, \mathrm{eV}^2
\le \Delta m^2 \le
10^{3} \, \mathrm{eV}^2 $.}
\label{aelamu}
\end{figure}

\begin{figure}[h]
\protect\caption{Exclusion regions in the
$ A_{\nu_{\mu};\nu_{e}} $--$ \Delta m^2 $
plane
for small
$ \left| U_{e3} \right|^2 $
and
$ \left| U_{\mu3} \right|^2 $.
The regions excluded by
the BNL E776 and KARMEN
$ \nu_\mu \leftrightarrows \nu_e $
appearance experiments
are bounded by the dash-dotted and dash-dot-dotted curves,
respectively.
The dashed line represents
the results of the Bugey experiment.
The curve passing through the circles
is obtained
from the results
of the Bugey, CDHS and CCFR84 experiments
using Eq.(\protect\ref{601}).
The curve passing through the triangles
is obtained
from the results
of the Bugey, FNAL E531 and CCFR95 experiments
using Eq.(\protect\ref{725}).
The dotted line is obtained
from the results
of the FNAL E531 and CCFR95 experiments
using Eq.(\protect\ref{825}).
The line passing through the squares
bounds the region that will be explored by
CHORUS and NOMAD.
The region allowed by the LSND experiment
is also shown
(the shadowed region limited by the two solid curves).}
\label{r1amuel}
\end{figure}

\begin{figure}[h]
\protect\caption{Exclusion regions in the
$ A_{\nu_{e};\nu_{\tau}} $--$ \Delta m^2 $
plane
for small
$ \left| U_{e3} \right|^2 $
and large
$ \left| U_{\mu3} \right|^2 $.
The solid line represents
the results of the FNAL E531 experiment.
The dashed line was obtained from
the results of the Bugey experiment.
The curve passing through the triangles
is obtained
from the results
of the Bugey, CDHS and CCFR84 experiments
using Eq.(\protect\ref{404}).
The curve passing through the squares
is obtained
from the results
of the CDHS, CCFR84 and BNL E776 experiments
using Eq.(\protect\ref{425}).
The curve passing through the circles
is obtained
from the results
of the BNL E776, FNAL E531 and CCFR95 experiments
using Eq.(\protect\ref{502}).}
\label{r2aelta}
\end{figure}

\begin{figure}[h]
\protect\caption{Upper bounds for
$ \left| U_{e3} \right|^2 $
in the region of small
$ \left| U_{e3} \right|^2 $
and large
$ \left| U_{\mu3} \right|^2 $.
The dash-dotted and dash-dot-dotted curves
are obtained from the results of the Bugey experiment
and
from the results of the CDHS and CCFR84 experiments,
respectively
(with the help of Eq.(\protect\ref{444})).
The dashed curve
is obtained from the results of the BNL E776 experiment
using Eq.(\protect\ref{410}).
The shadowed region within the two solid lines
is the allowed region obtained
from the results of the LSND experiment
using Eq.(\protect\ref{411}).}
\label{r2uel}
\end{figure}

\begin{figure}[h]
\protect\caption{Upper bounds for
$ 1 - \left| U_{\mu3} \right|^2 $
in the region of small
$ \left| U_{e3} \right|^2 $
and large
$ \left| U_{\mu3} \right|^2 $.
The solid curve
was obtained from the results of the CDHS and CCFR95 experiments
(see Eq.(\protect\ref{400})).
The dashed curve
was obtained from the results
of the BNL E776, FNAL E531 and CCFR95 experiments
using Eq.(\protect\ref{501}).}
\label{r2umu}
\end{figure}

\begin{figure}[h]
\protect\caption{Boundary curve in the
$ \Delta m^2 $--$ \left| \langle m \rangle \right| $
plane
for small
$ \left| U_{e3} \right|^2 $
and
$ \left| U_{\mu3} \right|^2 $.
The curve
was obtained from the results of the Bugey experiment.}
\label{r1dbeta}
\end{figure}

\begin{figure}[h]
\protect\caption{Boundary curves in the
$ \Delta m^2 $--$ \left| \langle m \rangle \right| $
plane
for small
$ \left| U_{e3} \right|^2 $
and large
$ \left| U_{\mu3} \right|^2 $.
The dash-dotted and dash-dot-dotted curves
are obtained from the results of the Bugey experiment
and
the results of the CDHS and CCFR84 experiments,
respectively.
The dashed curve
is obtained from the results of the BNL E776 experiment.
The shadowed region within the two solid lines
is the allowed region obtained
from the results of the LSND experiment.}
\label{r2dbeta}
\end{figure}

\begin{figure}[h]
\protect\caption{The regions
\protect\ref{R1} and \protect\ref{R2}
of the values of the parameters
$ \left| U_{e3} \right|^2 $
and
$ \left| U_{\mu3} \right|^2 $
for
$ \Delta m^2 = 6 \, \mathrm{eV}^2 $.
The vertical axis has been expanded logarithmically
for $ \left| U_{\mu3} \right|^2 $
very small and very large (i.e. close to one)
using the two different coordinates,
$ \left| U_{\mu3} \right|^2 $
below the horizontal solid line
and
$ ( 1 - \left| U_{\mu3} \right|^2 ) $
above the horizontal solid line.
We have drawn the constraints given
by the results of the
Bugey and CDHS
disappearance
experiments,
which delimit the allowed regions
\protect\ref{R1} and \protect\ref{R2}.
The black region is excluded by unitarity.}
\label{uelumu1}
\end{figure}

\begin{figure}[h]
\protect\caption{Allowed regions
for the values of the parameters
$ \left| U_{e3} \right|^2 $
and
$ \left| U_{\mu3} \right|^2 $
for
$ \Delta m^2 = 6 \, \mathrm{eV}^2 $.
The vertical axis has been expanded logarithmically
as in Fig.\protect\ref{uelumu1}.
We have drawn the constraints given
by the results of the
Bugey and CDHS
disappearance experiments
(as in Fig.\protect\ref{uelumu1})
and the constraints given
by the results of the
BNL E776 ($\nu_\mu\to\nu_e$)
and
FNAL E531 ($\nu_\mu\to\nu_\tau$)
experiments.
The region allowed by the LSND experiment
is shown as a lightly shadowed band.
The darkly shadowed region is excluded by unitarity.
The region Ia
is the part of region \protect\ref{R1}
which is allowed by all experiments,
except LSND.
The region IIa is the part of
region \protect\ref{R2} which is allowed by all experiments,
except LSND,
whereas the region IIb is the part of
region \protect\ref{R2} which is allowed by all experiments,
including LSND.}
\label{uelumu2}
\end{figure}

%\end{document}
%\pagestyle{empty}

\newpage

\begin{minipage}[h]{\textwidth}
\null\vskip-1cm
\begin{center}
\mbox{\epsfig{file=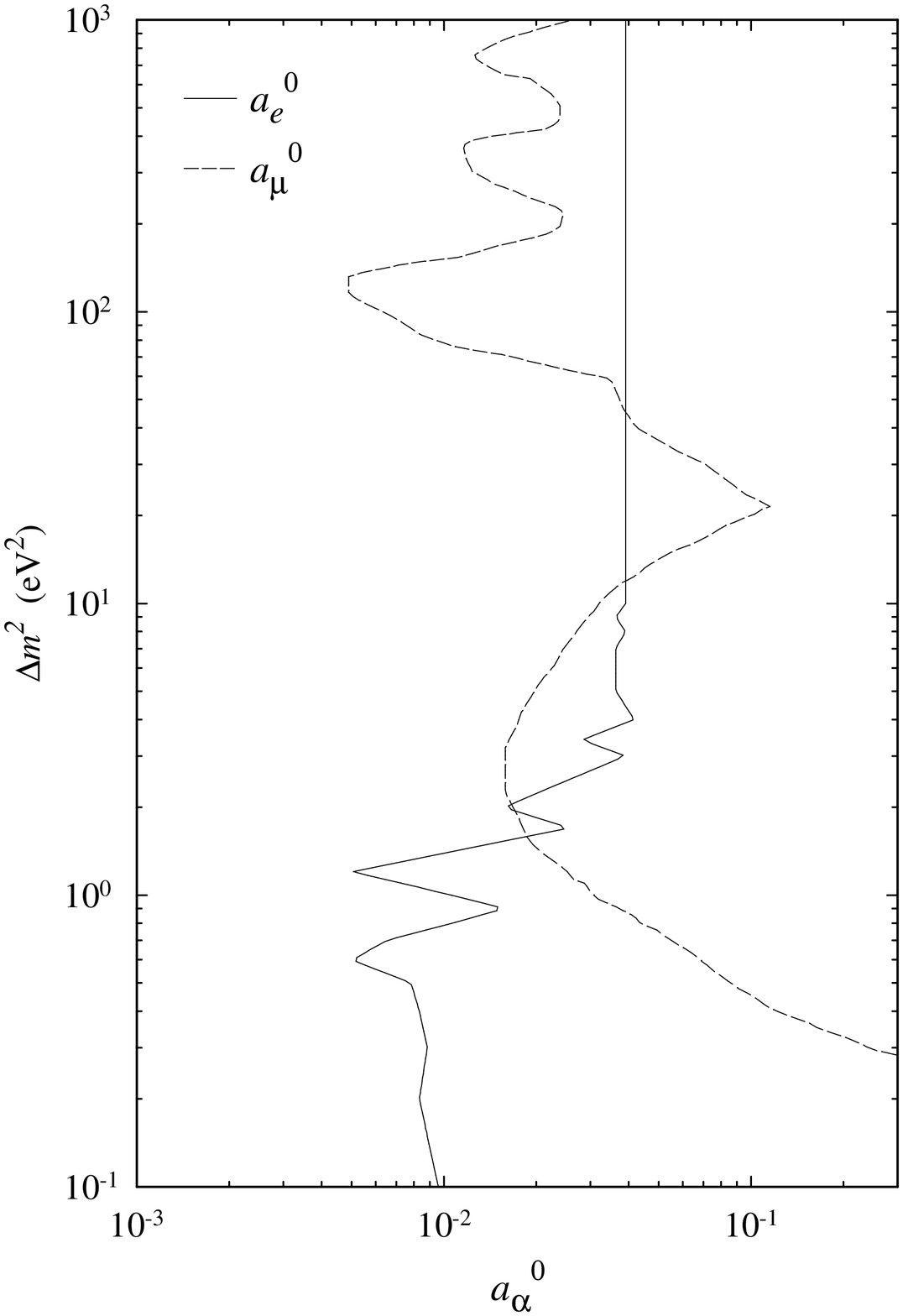,height=0.9\textheight}}
\end{center}
\end{minipage}
\vspace{1cm}
\begin{center}
{\Large Figure \ref{aelamu}}
\end{center}

\newpage

\begin{minipage}[h]{\textwidth}
\null\vskip-1cm
\begin{center}
\mbox{\epsfig{file=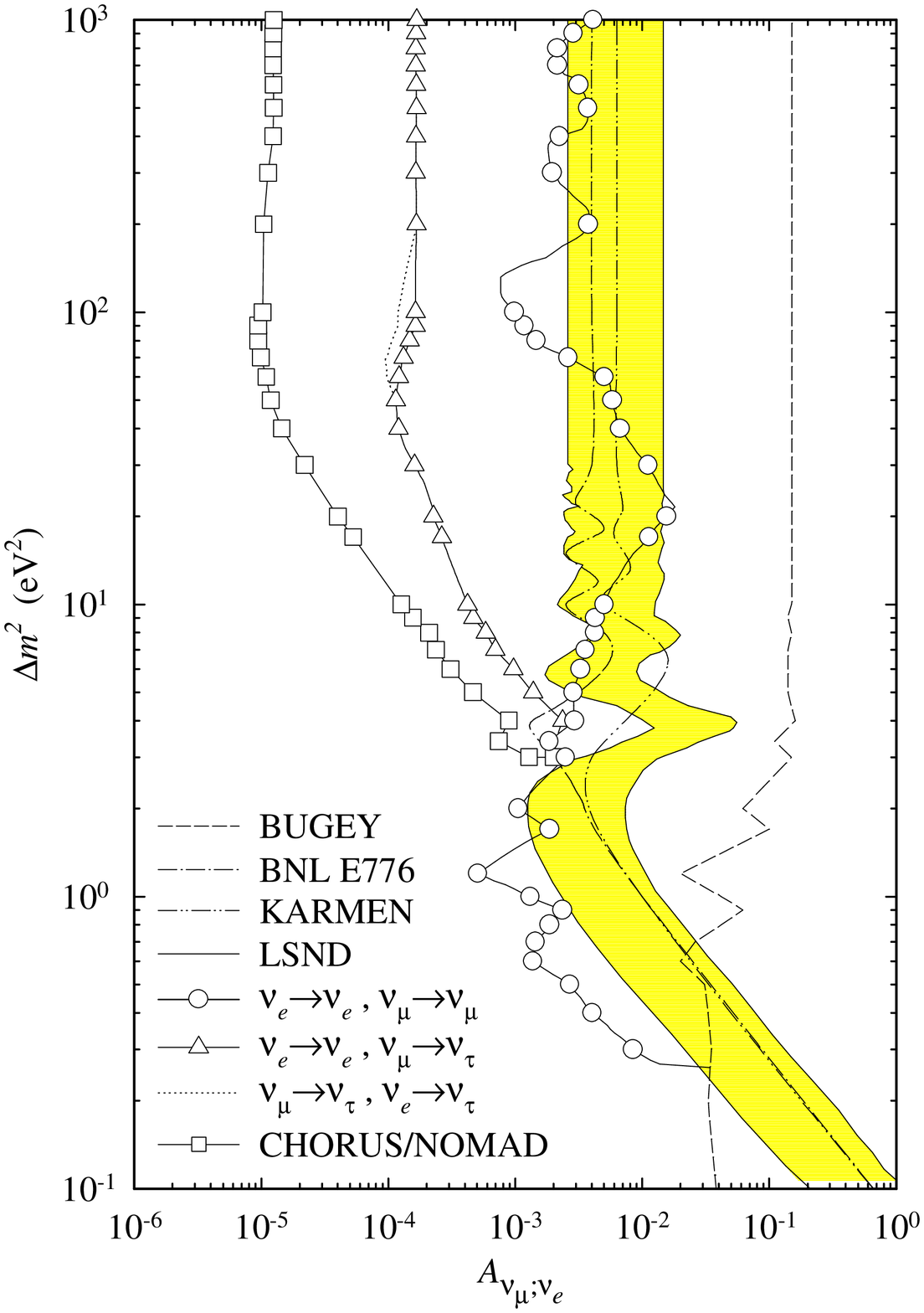,height=0.9\textheight}}
\end{center}
\end{minipage}
\vspace{1cm}
\begin{center}
{\Large Figure \ref{r1amuel}}
\end{center}

\newpage

\begin{minipage}[h]{\textwidth}
\null\vskip-1cm
\begin{center}
\mbox{\epsfig{file=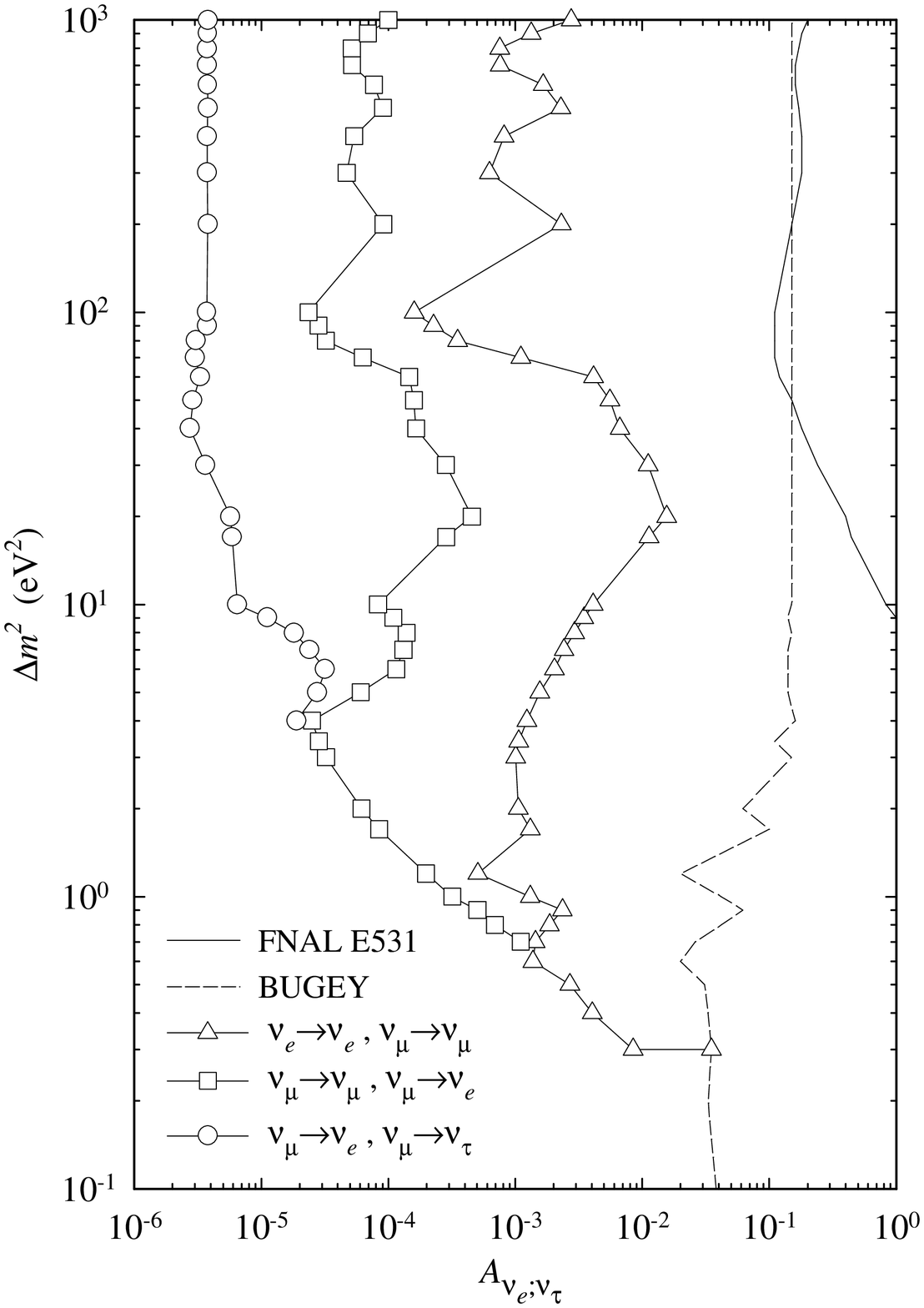,height=0.9\textheight}}
\end{center}
\end{minipage}
\vspace{1cm}
\begin{center}
{\Large Figure \ref{r2aelta}}
\end{center}

\newpage

\begin{minipage}[h]{\textwidth}
\null\vskip-1cm
\begin{center}
\mbox{\epsfig{file=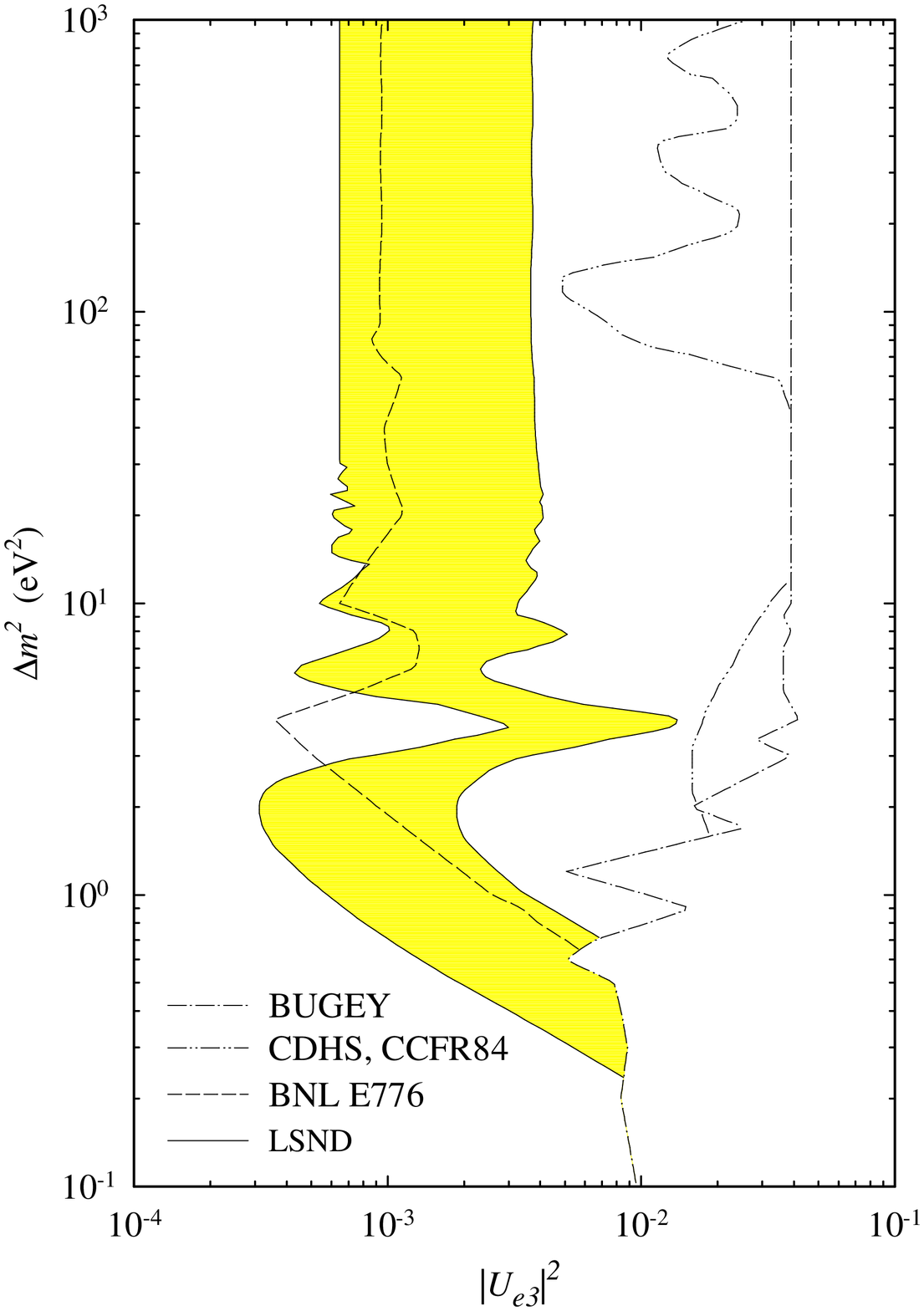,height=0.9\textheight}}
\end{center}
\end{minipage}
\vspace{1cm}
\begin{center}
{\Large Figure \ref{r2uel}}
\end{center}

\newpage

\begin{minipage}[h]{\textwidth}
\null\vskip-1cm
\begin{center}
\mbox{\epsfig{file=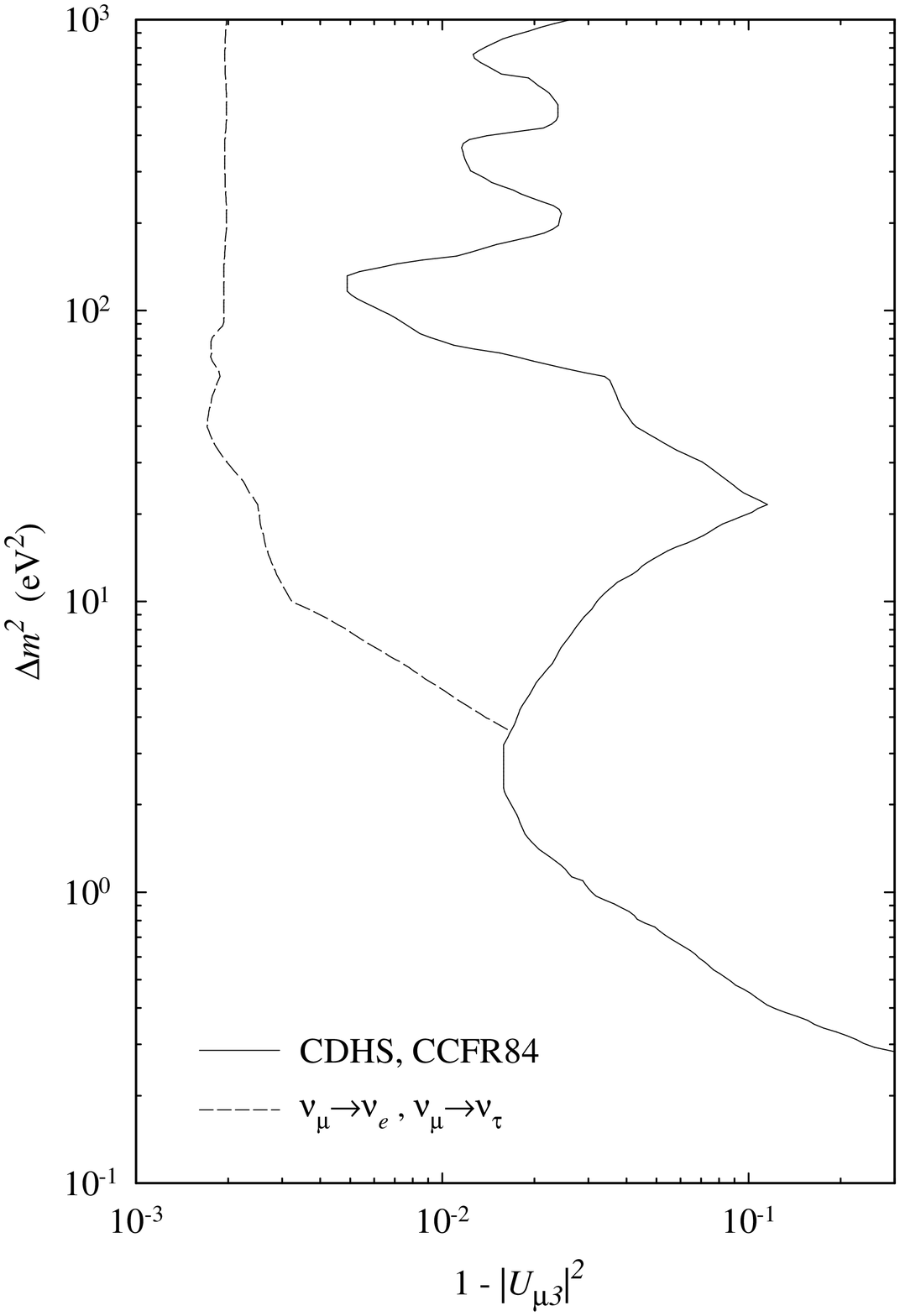,height=0.9\textheight}}
\end{center}
\end{minipage}
\vspace{1cm}
\begin{center}
{\Large Figure \ref{r2umu}}
\end{center}

\newpage

\begin{minipage}[h]{\textwidth}
\null\vskip-1cm
\begin{center}
\mbox{\epsfig{file=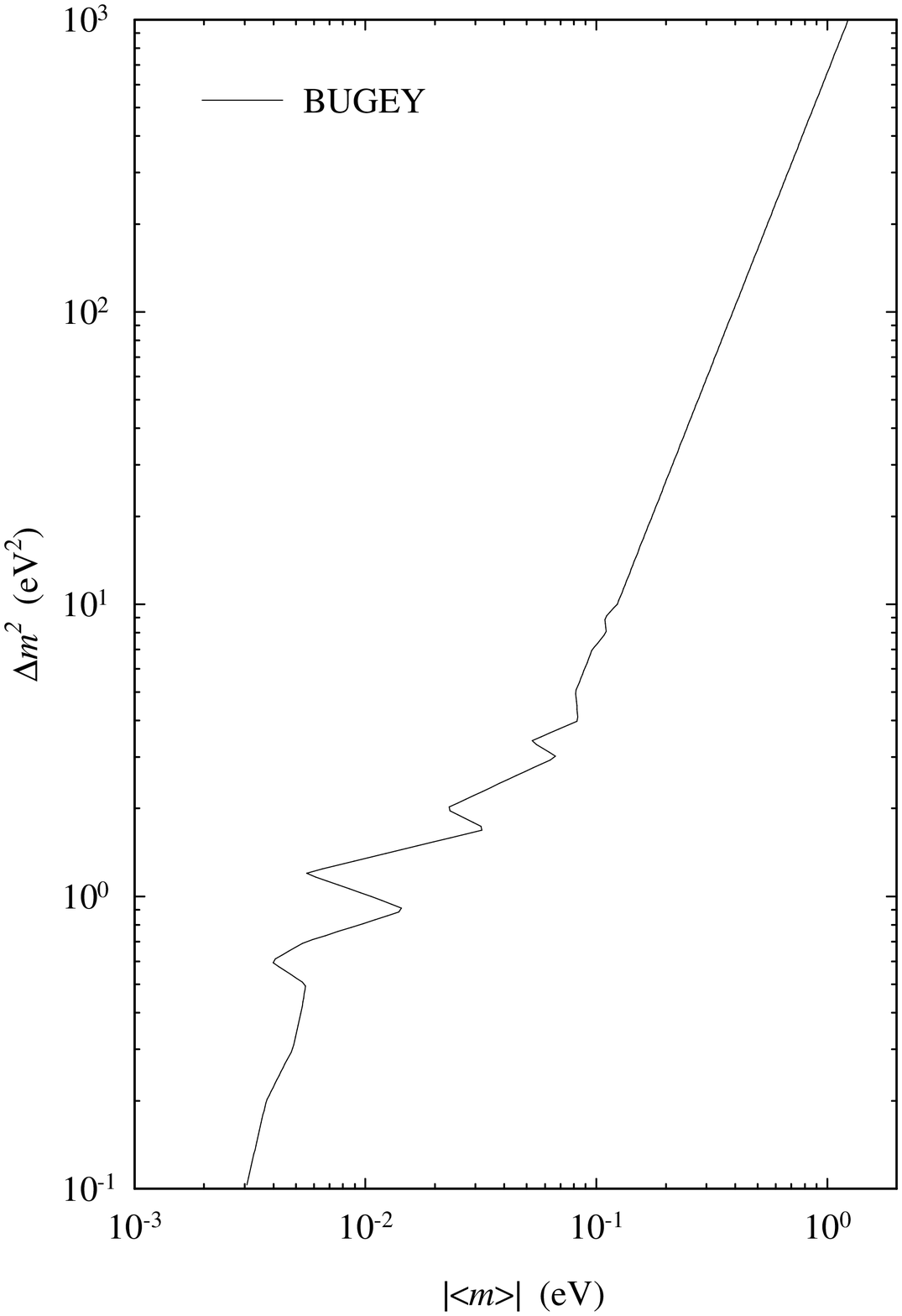,height=0.9\textheight}}
\end{center}
\end{minipage}
\vspace{1cm}
\begin{center}
{\Large Figure \ref{r1dbeta}}
\end{center}

\newpage

\begin{minipage}[h]{\textwidth}
\null\vskip-1cm
\begin{center}
\mbox{\epsfig{file=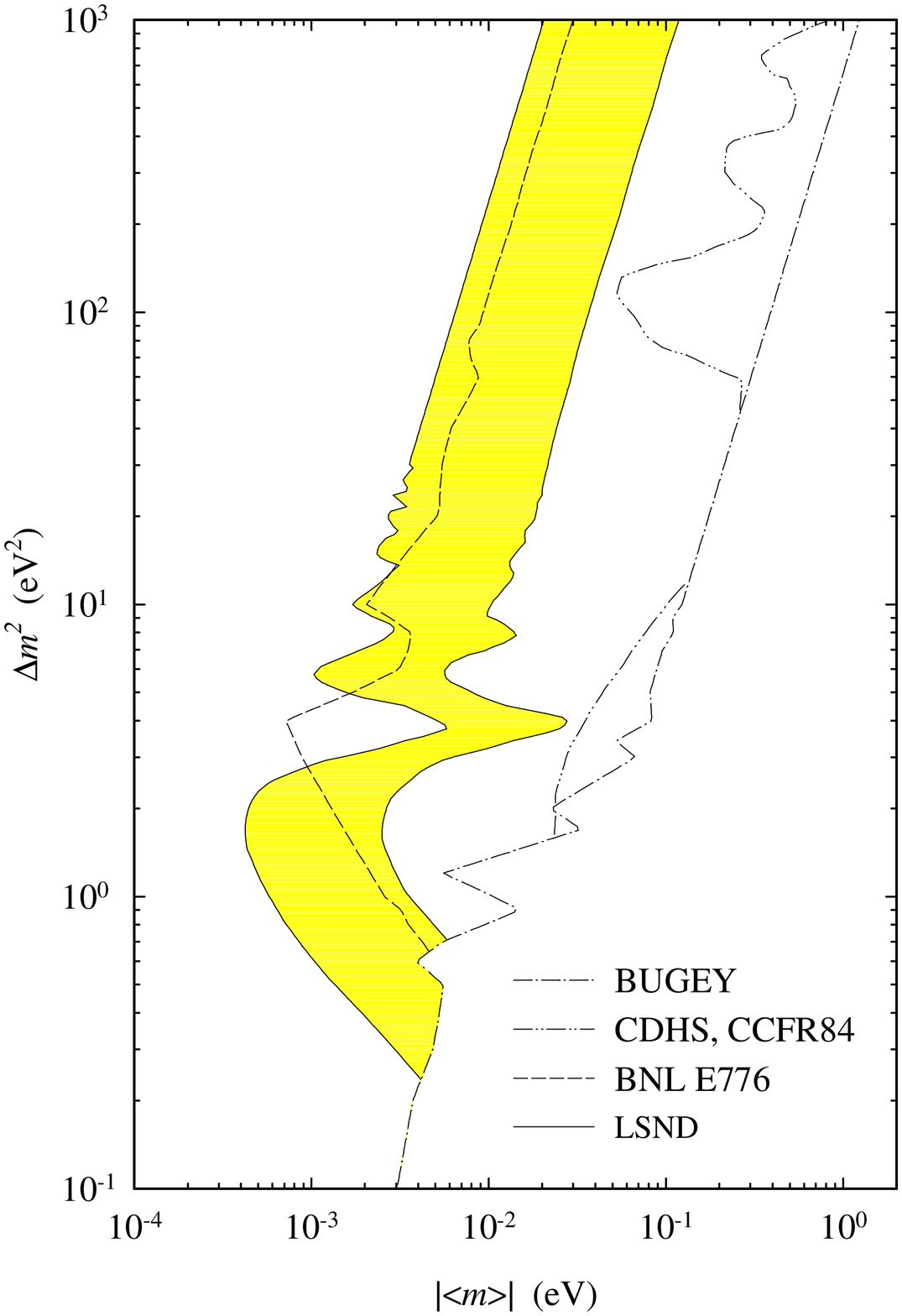,height=0.9\textheight}}
\end{center}
\end{minipage}
\vspace{1cm}
\begin{center}
{\Large Figure \ref{r2dbeta}}
\end{center}

\newpage

\begin{minipage}[h]{\textwidth}
\null\vskip-1cm
\begin{center}
\mbox{\epsfig{file=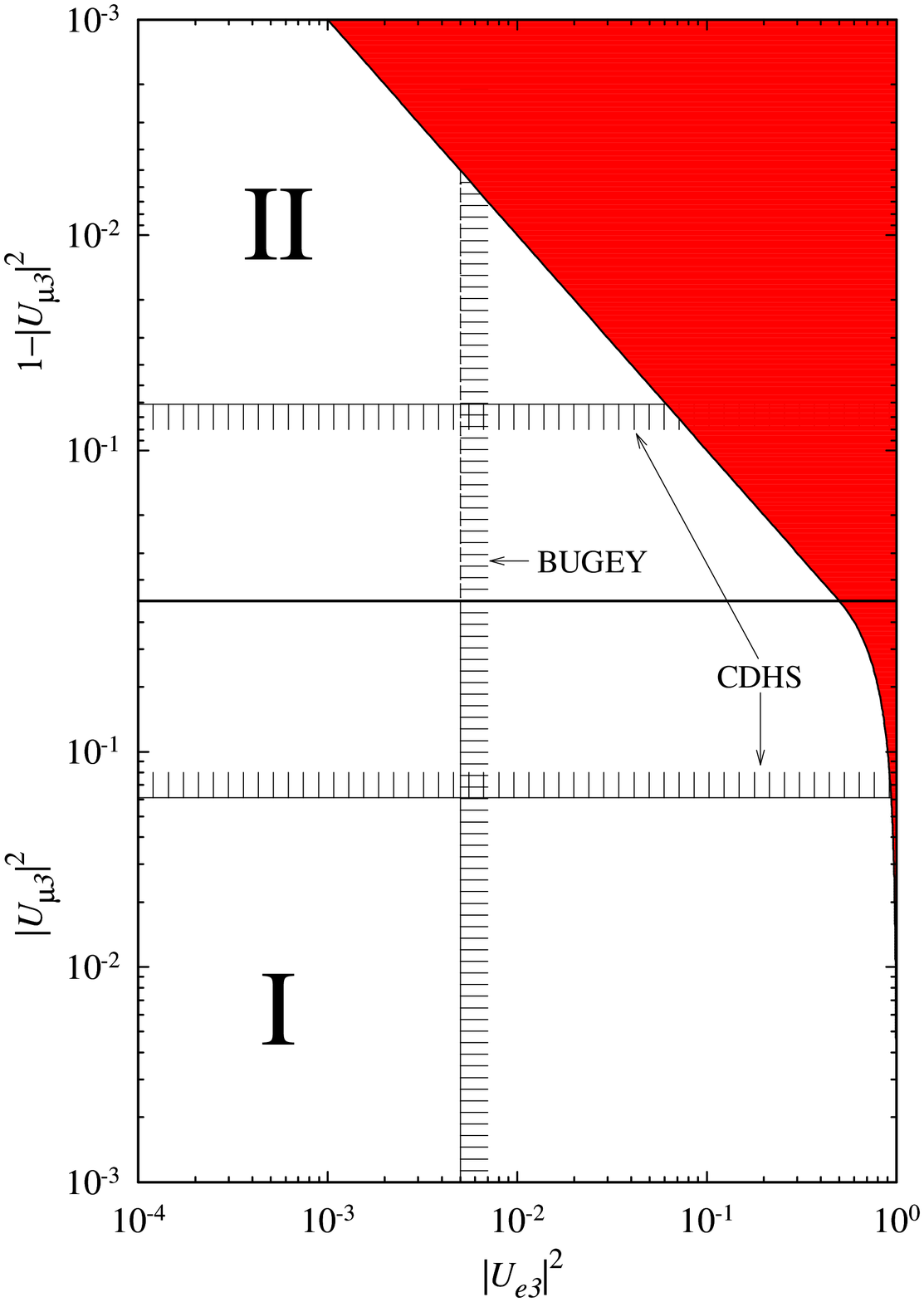,height=0.9\textheight}}
\end{center}
\end{minipage}
\vspace{1cm}
\begin{center}
{\Large Figure \ref{uelumu1}}
\end{center}

\newpage

\begin{minipage}[h]{\textwidth}
\null\vskip-1cm
\begin{center}
\mbox{\epsfig{file=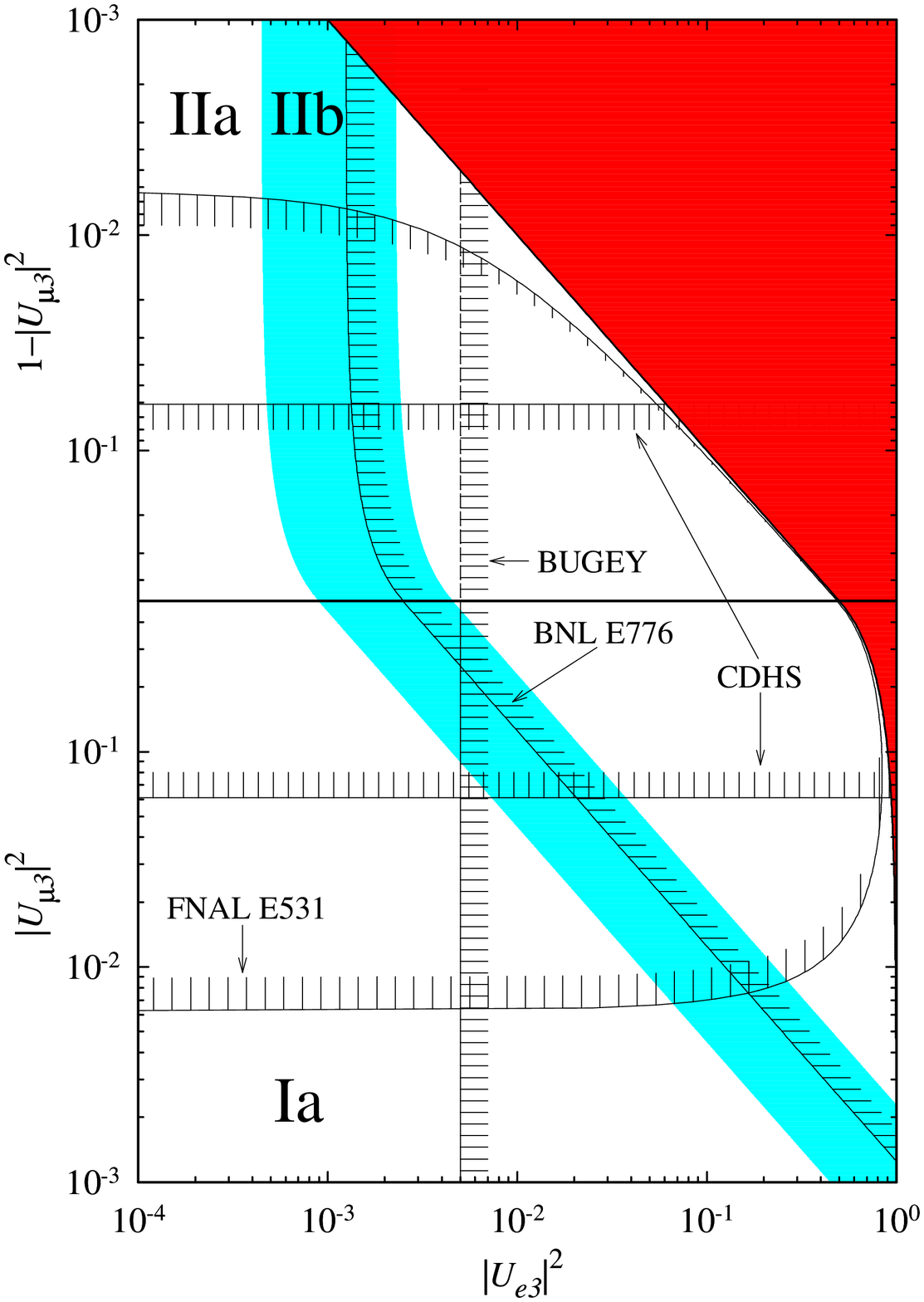,height=0.9\textheight}}
\end{center}
\end{minipage}
\vspace{1cm}
\begin{center}
{\Large Figure \ref{uelumu2}}
\end{center}

\end{document}